\documentclass[%
reprint, 
superscriptaddress,
longbibliography
amsmath,amssymb,
aps, physrev,
]{revtex4-2}

\usepackage{graphicx}
\usepackage{dcolumn}
\usepackage{xcolor}
\usepackage{bm}
\usepackage{float}
\makeatletter
\let\newfloat\newfloat@ltx
\makeatother
\usepackage{algorithm}
\usepackage{algpseudocode}
\usepackage{placeins}
\usepackage{amsmath}
\usepackage{textgreek}
\usepackage[colorlinks=true,breaklinks=true,linkcolor=magenta,citecolor=magenta,urlcolor=magenta]{hyperref}
\newcommand{\scientific}[2]{$#1 \cdot 10^{#2}$}

\begin{document}

\preprint{APS/123-QED}

\title{Sample-based quantum diagonalization as parallel fragment solver for the localized active space self-consistent field method}

\author{Qiaohong Wang}
 \affiliation{Pritzker School of Molecular Engineering, University of Chicago, Chicago, IL 60637, USA}
\author{Mario Motta}
\affiliation{IBM Quantum, IBM T.J. Watson Research Center, Yorktown Heights, NY 10598, USA}
\author{Ruhee D'Cunha}\thanks{Currently at Millikin University, Decatur, IL 62522, USA.}
\affiliation{Department of Chemistry, Chicago Center for Theoretical Chemistry, University of Chicago, Chicago, IL 60637, USA}
\author{Kevin J. Sung}
\affiliation{IBM Quantum, IBM T.J. Watson Research Center, Yorktown Heights, NY 10598, USA}
\author{Matthew R. Hermes}
\affiliation{Department of Chemistry, Chicago Center for Theoretical Chemistry, University of Chicago, Chicago, IL 60637, USA}
\author{Tanvi Gujarati}
\affiliation{IBM Quantum, IBM Research -- Almaden, San Jose, CA 95120, USA}
\author{Yukio Kawashima}
\affiliation{IBM Quantum, IBM Research -- Tokyo, Tokyo 103-8510, Japan}
\author{Yu-ya Ohnishi}
\affiliation{Materials Informatics Initiative, RD technology and digital transformation center, JSR Corporation, Kawasaki-ku, Kawasaki, Kanagawa 210-0821, Japan}
\author{Gavin O. Jones}
\affiliation{IBM Quantum, IBM T.J. Watson Research Center, Yorktown Heights, NY 10598, USA}
\author{Laura Gagliardi}
\email{lgagliardi@uchicago.edu}
\affiliation{Pritzker School of Molecular Engineering, University of Chicago, Chicago, IL 60637, USA}
\affiliation{Department of Chemistry, Chicago Center for Theoretical Chemistry, University of Chicago, Chicago, IL 60637, USA}

\date{\today}

\begin{abstract}
Accurately and efficiently describing strongly correlated electronic systems is a central challenge in quantum computational chemistry, with classical and quantum computers. The localized active space self-consistent field method (LASSCF) uses a product of fragment active spaces as a variational space, with the Schrödinger equation solved exactly in each fragment and the fragment active-space orbitals defined in a self-consistent manner. LASSCF is accurate for systems with strong intra-fragment and weak inter-fragment correlation, and its computational cost is combinatorial with respect to the size of the individual fragment active spaces, rather than their product. However, exactly solving the Schrödinger equation in each fragment remains a substantial bottleneck. Here, we address the possibility of solving the fragment active space Schrödinger equation with approximate methods, particularly sample-based quantum diagonalization (SQD). SQD is a technique that uses a quantum computer to sample configurations from a chemically motivated quantum circuit and a classical computer to mitigate errors and solve the Schrödinger equation in a subspace of the configuration space. We apply the proposed method, LASSQD, to the [Fe(H$_2$O)$_4$]$_2$bpym$^{4+}$ compound and the [Fe$^{\mathrm{III}}$Fe$^{\mathrm{III}}$Fe$^{\mathrm{II}}$($\mu$$_3$-O)-(HCOO)$_6$] complex for calculating the intermediate-spin ground state energies, assessing its accuracy and precision, respectively originating from approximations in the solution of the Schrödinger equation and stochasticity of configuration sampling. We observe that LASSQD can tackle fragment sizes intractable by LASSCF, achieves within 1kcal/mol agreement to LASSCF, and delivers results that are competitive with alternative classical methods to solve the Schrödinger equation, and thus can be used as a starting point for a perturbative treatment (LASSQD-PDFT) to recover correlation external to the active space.

\end{abstract}

\maketitle


\section{\label{sec:level1}Introduction}
Exact diagonalization or full configuration interaction (FCI) provides numerically exact solutions to the electronic, time-independent Schrödinger equation under the Born-Oppenheimer approximation, for a given finite orbital basis. However, the computational cost of FCI scales combinatorially with the number of electrons and orbitals, limiting its practicality for large systems. To date, the largest reported demonstration involves 26 electrons in 23 orbitals~\cite{2623}. To solve larger systems, especially molecules with multiple transition metals, methods such as the complete active space self-consistent field (CASSCF)~\cite{casscf} restrict FCI to a subset of chemically relevant orbitals. However, 
the computational cost of CASSCF still scales combinatorially with the size of the active space (AS), which severely limits its practicality~\cite{22}.

One direction to reduce the cost of CASSCF involves fragmentation based on chemical intuition, for example, the generalized active space self-consistent field (GASSCF)~\cite{gasscf1,gasscf2,gasscf3,gasscf4}, restricted active space self-consistent field (RASSCF)~\cite{rasscf,rasscf2}, and localized active space self-consistent field (LASSCF) method~\cite{LAS,vLASSCF}, also known as the cluster mean-field (cMF) method~\cite{Jimenez-Hoyos2015}. LASSCF approximates the active-space wave function as a single antisymmetrized product of smaller active-space wave functions, with the assumption that such active spaces are relatively weakly interacting. Previously, LASSCF has been shown to accurately describe the dissociation of two double bonds in bisdiazene~\cite{vLASSCF}, to generate more consistent orbital evolutions throughout the potential energy surface of sulfonium salts compared to CASSCF~\cite{wangsalts}, and 
to predict spin-state energy gaps close to those of CASSCF for bimetallic compounds~\cite{riddhish_spin}. However, the computational cost of LASSCF within the fragments still scales exponentially, thus limiting the size of the fragments and the resulting active spaces that one can study. 

Another direction to reduce the cost of CASSCF 
involves approximately solving the active-space Schrödinger equation. Efforts to approximate CASSCF active space FCI solvers include density matrix renormalization group~\cite{dmrgscf1,dmrgscf2,dmrgscf3}, multiple flavors of selected configuration interaction (SCI) followed by perturbation theory~\cite{hciscf,asci_scf,ice_scf_pt2,asci_scf_pt2}, variational two-electron reduced-density-matrix (v2RDM)~\cite{v2rdm_cas} and semistochastic/stochastic methods like Full CI quantum Monte Carlo (FCIQMC)~\cite{fciqmc_scf1,fciqmc_scf2,fciqmc_scf_is}, among others. These approaches aim to accurately approximate the exact solution of the active-space (or fragment) Schrödinger equation with a significantly lower computational cost than exact diagonalization.

Recent advances in quantum computation have produced ground-state approximation methods that can be applied within a limited computational budget on today's hardware.
Examples include the variational quantum eigensolver (VQE)~\cite{Kandala2017,VQE,VQE_review,McClean_2016}, quantum selected configuration interaction (QSCI)~\cite{qsci_og,adapt_qsci}, and sample-based quantum diagonalization (SQD)~\cite{SQD}. SQD has drastically increased the possibility of running quantum chemical algorithms on near-term devices, as well as the number of qubits used in quantum chemical calculations, with the largest demonstration to date, of seventy-seven qubits for a single point energy calculation of the [Fe$_4$S$_4$] cluster. 


In this work, we introduce LASSQD, a method that uses SQD as the fragment solver to approximate LASSCF. It is the first time an approximate solver is used in the fragments, which allows us to assess the impact of approximations in the fragment active space solution and solve fragment sizes intractable by FCI in the LASSCF framework. We successfully mitigate a challenge of the SQD procedure when combined with SCF orbital optimization, because of the sampling stochasticity, by introducing a retention step. Lastly, we also implement pair-density functional theory~\cite{mcpdft}, LASSQD-PDFT, to include correlations external to the active space, making LASSQD-like methods a competitive candidate for challenging metallic systems. By leveraging SQD, we can probe how near-term algorithms on current quantum devices may serve as scalable active space solvers for multireference quantum chemistry. LASSQD provides a framework for assessing the current limitations and future potential of quantum hardware and software, enabling systematic progress toward solving challenging multireference systems on quantum devices.

The rest of the paper is laid out as follows: in Section II, we introduce key methods, such as LASSCF, SQD, carryover SQD, local unitary cluster Jastrow (LUCJ) ansatz, and the implementation of LASSQD, as well as LASSQD-PDFT. In Section III, we first present results of LASSQD when SQD is the same as FCI in LAS fragments of bimetallic compound [Fe(H$_2$O)$_4$]$_2$bpym$^{4+}$. Then, we discuss the SQD approximation of FCI in LAS fragments, specifically the stochasticity challenge and carryover LASSQD. We find that with carryover, LASSQD efficiently approximates the FCI space while allowing for meaningful orbital optimization of both bimetallic compound [Fe(H$_2$O)$_4$]$_2$bpym$^{4+}$ and trimetallic compound [Fe$^{\mathrm{III}}$Fe$^{\mathrm{III}}$Fe$^{\mathrm{II}}$($\mu$$_3$-O)-(HCOO)$_6$]. Lastly, we demonstrate LASSQD calculations on iron-porphyrin, which has one fragment size intractable for classical FCI. We test the limits of current hardware and of quantum methods for efficiently solving fragments of such sizes. In Section IV, we discuss future directions.

\section{Methods\label{SecII}}
\subsection{LASSCF}
In the LASSCF framework, the total energy is obtained by variationally minimizing the expectation value of the second quantized molecular Hamiltonian, 
\begin{equation}
    \hat{H}=h^p_q\hat{a}_p^\dagger \hat{a}_q + \frac{1}{4}h_{qs}^{pr}\hat{a}_p^\dagger \hat{a}_r^\dagger \hat{a}_s\hat{a}_q
\end{equation}
with respect to the fragment active space configuration coefficients $\mathbf{x_{\vec{a}_i}}$, and orbital rotations for the full set of of molecular orbitals $\mathbf{x_{r}^{q}}$, 
\begin{equation} 
E_{\mathrm{LASSCF}}=\min_{\mathbf{x_{r}^{q}},\mathbf{x_{\vec{a}_i}}}\langle \Psi_{\mathrm{LASSCF}}| \hat{H} | \Psi_{\mathrm{LASSCF}}\rangle. 
\end{equation} 
Where $h^p_q$ and $h_{qs}^{pr}$ are the one-electron and two-electron Hamiltonian matrix elements, $\hat{a}_p^\dagger$($\hat{a}_p$) creates (annihilates) an electron in spin-orbital $p$. $q,r$ index individual spin-orbitals in two different subspaces, and $\vec{a}_i$ is a determinant or configuration state function.   The ansatz for the wave function is expressed as \begin{equation}\label{equ:vlasscf}
\left|\Psi_{\mathrm{LASSCF}}\right\rangle=\left[\prod_K \hat{U}_{\mathrm{CI}, K}\right] \hat{U}_{\mathrm{orb}}| (\bigwedge_K \Psi_{A_K}) \wedge \Phi_D\rangle, 
\end{equation} 
where  $\Psi_{A_K}$ represents a correlated wave function localized on an active space labeled by an index $K$, and henceforth $\Psi_{A_K}$ is denoted as ``fragment" wave function. And $\Phi_D$ denotes a single determinant spanning the inactive or core space. The wedge symbol $\bigwedge$ indicates that the wave function is constructed as an antisymmetrized product of the fragment-local wave functions and the core determinant. The unitary operators that transform the LASSCF wave function are defined as:
\begin{equation}
        \hat{U}_{\text {orb }}=\exp \left[\sum_{q>r} \sum_r \mathbf{x}_r^q\left(\hat{a}_{q \sigma}^{\dagger} \hat{a}_{r \sigma}-\hat{a}_{r \sigma}^{\dagger} \hat{a}_{q \sigma}\right)\right],
\end{equation}
\begin{equation}
       \hat{U}_{\mathrm{CI}, K}=\exp \left[\sum_{\vec{a}_i} \mathbf{x}_{\vec{a}_i}\left(\left|\vec{a}_i\right\rangle\left\langle\Psi_K\right|-\left|\Psi_K\right\rangle\left\langle\vec{a}_i\right|\right)\right]. 
\end{equation}
In practice, the choice of fragments and their active spaces is made to capture the short-range static correlation within subunits of the system. Within the standard LASSCF procedure, each fragment's correlated wave function $\Psi_{A_K}$ is obtained by solving an FCI problem in that fragment's active orbitals, with the mean-field effect of the remaining fragments included self-consistently. The description of the specific LASSCF implementation used in this work is provided in Sec.~\ref{las_rdm}. 

\subsection{Quantum Algorithms}

\subsubsection{QSCI and SQD}
QSCI diagonalizes an effective Hamiltonian to obtain an approximate eigenvalue (i.e., ground-state energy) in a subspace of electronic configurations \textit{sampled} by a quantum computer. In other words the configuration selection is not an iterative and deterministic procedure as in SCI, but a random procedure using a quantum device to draw samples from a probability distribution. The QSCI framework has been further extended with numerous improvements to be compatible with near-term hardware limitations, including features of noise-resilient and shot-budget friendly, as sample-based quantum diagonalization (SQD)~\cite{SQD}. 

SQD has emerged as a promising classical-quantum hybrid algorithm to approximate the eigenstates of many-body systems. Here we present a short review of the algorithm. SQD's foundation lies in the nature of the quantum measurement, where sampling a quantum state $\Psi$ in the computational basis yields measurement outcomes in the form of bitstrings $\mathbf{x} \in \{0,1\}^{2N_{\mathrm{MO}}}$, where $N_\mathrm{MO}$ is the number of molecular orbitals. Each bitstring corresponds to a specific electronic configuration (i.e., the qubit basis states $|0\rangle$ ($|1\rangle$)) corresponds to empty (occupied) spin-orbitals under the Jordan-Wigner (JW) encoding~\cite{JW}. Repeating the above process generates a set of measurement outcomes:
\begin{equation}
\tilde{\mathcal{X}}=\left\{\mathbf{x} \mid \mathbf{x} \sim \widetilde{P}_{\Psi}(\mathbf{x})\right\}
\end{equation}
where the $\widetilde{P}_{\Psi}(\mathbf{x})$ dictates the probability of sampling that particular bitstring with the presence of noise. To parallelize the computational cost of the classical post-processing, the raw output space $\tilde{\mathcal{X}}$ is then subsampled into $K$ batches of configurations $\mathcal{S}^{(k)}, k = 1,..., K$ according to the same probability $\widetilde{P}_{\Psi}$ (i.e. empirical frequencies of each $\mathbf{x}$ in $\tilde{\mathcal{X}}$), of a fixed batch size $d$ (the dimension of the subspace). The many-body Hamiltonian can then be projected into this subspace as 
\begin{equation}\label{equ-4}
\hat{H}_{\mathcal{S}^{(k)}}=\hat{P}_{\mathcal{S}^{(k)}} \hat{H} \hat{P}_{\mathcal{S}^{(k)}}, \text { with } \hat{P}_{\mathcal{S}^{(k)}}=\sum_{\mathbf{x} \in \mathcal{S}^{(k)}}|\mathbf{x}\rangle\langle\mathbf{x}|,
\end{equation}
where we diagonalize the projected Hamiltonian 
\begin{equation}
\hat{H}_{\mathcal{S}^{(k)}}\left|\psi^{(k)}\right\rangle=E^{(k)}\left|\psi^{(k)}\right\rangle
\end{equation}
by solving the eigenvalue equation with a spin-constrained eigenstate solver, where $\hat S^{2}$ is the total spin operator:
\begin{equation}
    \qquad
\hat S^{2}_{\mathcal S^{(k)}}=\hat P_{\mathcal S^{(k)}}\hat S^{2}\hat P_{\mathcal S^{(k)}},
\end{equation}
\begin{equation}
\left(\hat{H}_{\mathcal{S}^{(k)}}+\lambda\left[\hat S^{2}_{\mathcal S^{(k)}}-s(s+1)\right]^2\right)|\psi^{(k)}\rangle=E^{(k)}|\psi^{(k)}\rangle.       
\end{equation}
Here, $\lambda$ is the Lagrange multiplier, which is set to a default value of $\lambda =0.2$.

In the first iteration, bitstrings sampled from quantum devices with incorrect particle number and $z$-component of the spin are temporarily excluded from the electronic configurations in $\tilde{\mathcal{X}}$. The subsequent batches $\mathcal{S}^{(k)}$ are constructed, onto which the Hamiltonian $\hat{H}_{\mathcal{S}^{(k)}}$ is projected and diagonalized, as eq.\ref{equ-4} shows. From each batch and therefore each diagonalization, we obtain an expected number of electrons in each spin orbital. Averaging these occupancies across batches yields an average occupation number for each spin orbital
\begin{equation}
    n_{p\sigma} = \frac{1}{K}\sum_{1\leq k\leq K}\langle \psi^{(k)}|\hat{a}^{\dagger}_{p\sigma}\hat{a}_{p\sigma}|\psi^{(k)}\rangle,
\end{equation}
where $p = 0,1,...N_{\mathrm{MO}}-1$, $N_{\mathrm{MO}}$ is the number of orbitals and $\sigma \in {\alpha, \beta}$ spin polarizations. 
In subsequent iterations, a self-consistent loop restores particle number conservation using $n_{p\sigma}$. First, bitstrings $\textbf{x}$ with the incorrect particle number and $z$-component of the spin (i.e., $N_\textbf{x} \neq N$ particles) undergo a bit-flip modification to be in the right symmetry. Specifically, if $N_\textbf{x} > N$ or $N_\textbf{x} < N$, $|N_\textbf{x}-N|$ bits are further sampled to be flipped from the set of occupied or empty spin orbitals, with selection probabilities weighted by the distance $|x_{p\sigma}-n_{p\sigma}|$ between the bit value and the current average orbital occupancy obtained from the previous round for each spin orbital. This procedure generates a new set of recovered configurations $\mathcal{X}_R$. Second, the set of recovered bitstrings goes through the regular procedure of constructing batches $\mathcal{S}^{(k)}$, onto which the Hamiltonian $\hat{H}_{\mathcal{S}^{(k)}}$ is further projected, and diagonalized in each batch. The target eigenstate $|\psi^{(k)}\rangle$ is chosen as the lowest-energy one among the batches. Then, the average orbital occupancy is computed again for the next iteration's bitstring correction. The SQD procedure is illustrated in Algorithm \ref{alg:sqd}.

\begin{algorithm}
\caption{Sample-based Quantum Diagonalization (SQD)\label{alg:sqd}}
\begin{algorithmic}[1]
\State \textbf{Input:} Bitstrings $\mathbf{x}$, Hamiltonian $\hat{H}$, number of batches $k$, samples per batch $d$, number of iterations $N_{\mathrm{iter}}$, spin and particle number targets
\State \textbf{Output:} Approximate eigenvalue $E^{(k)}$, eigenstate $|\psi^{(k)}\rangle$

\For{iteration $= 1$ to $N_{\mathrm{iter}}$}
    \If{iteration $= 1$}
        \State Discard bitstrings with incorrect particle number 
    \Else
        \State Modify bitstrings via bit-flip correction using previous $n_{p\sigma}$ to match target symmetry
    \EndIf
    \State Group valid bitstrings into batches: $\mathcal{S}^{(k)}$
    \For{each batch $\mathcal{S}^{(k)}$}
        \State Project the Hamiltonian: $\hat{H}_{\mathcal{S}^{(k)}} = \hat{P}_{\mathcal{S}^{(k)}} \hat{H} \hat{P}_{\mathcal{S}^{(k)}}$
        \State Solve eigenvalue equation with spin penalty:
        $\left(\hat{H}_{\mathcal{S}^{(k)}} + \lambda(\hat S^{2}_{\mathcal S^{(k)}} - s(s+1))^2 \right) |\psi^{(k)}\rangle = E^{(k)} |\psi^{(k)}\rangle$
        \State Compute and update spin-resolved orbital occupancy: 
        $n_{p\sigma} =\frac{1}{K}\sum_{1\leq k\leq K} \langle \psi^{(k)} | \hat{a}^\dagger_{p\sigma} \hat{a}_{p\sigma} | \psi^{(k)} \rangle$
    \EndFor
    \State Select the lowest-energy eigenstate from all batches
\EndFor
\State \Return $E^{(k)}, |\psi^{(k)}\rangle$
\end{algorithmic}
\end{algorithm}

\subsubsection{SQD with carryover}
At each iteration of configuration recovery, one discards information from previous iterations, so the energy may not converge and fluctuate considerably. To improve, we employ a revised version of SQD that incorporates a carryover mechanism~\cite{carroyver}, in which bitstrings from one iteration are retained and reintroduced into the subspace of subsequent iterations. By carrying over these configurations, SQD preserves the most significant correlation contributions. This targeted reuse of important configurations improves convergence efficiency by guiding the algorithm toward physically relevant electronic configurations in the iterative LASSQD procedure, which we will discuss the integration in the LAS structure in Section \ref{lassqd} and show results in Section \ref{carryoverlassqd}. The carryover SQD is describe below in Algorithm \ref{alg:co_sqd}.

In the first iteration, all steps are the same as conventional SQD, and the batch with the lowest energy eigenstate is identified. The 
CI-coefficient tensor of this solution is saved and used for the following analysis. First, the amplitudes of all determinants are collected into a one-dimensional array and sorted by absolute value. Then, determinants with amplitudes exceeding a user-defined threshold are selected. These selected determinants are subsequently mapped back into  $\alpha$ and $\beta$ strings, referred to as ``carryover" determinants.

In the subsequent iteration, in addition to the self-consistent loop, we append any saved bitstrings that are not already present in the current iteration and update after each iteration. Specifically, after constructing batches $\mathcal{S}^{(k)}$, we extend the batch with unique bitstrings from the carryover procedure,
\begin{equation}
    \mathcal{S}^{(k)} \cup (\mathcal{C}_\alpha\otimes\mathcal{C}_\beta)\to \tilde{\mathcal{S}}^{(k)}.
\end{equation}
The modified batches are further used for diagonalization.
Note that the size of the diagonalization subspace, $|\tilde{\mathcal{S}}^{(k)}|$, is now affected by both the input samples per batch $d$ and the carryover cutoff $\epsilon$. When all batches are diagonalized, we identify the lowest energy eigenstate again and repeat the above to save the new ``carryover" bitstrings.

\begin{algorithm}
\caption{Carryover SQD\label{alg:co_sqd}}
\begin{algorithmic}[1]
\State\textbf{Input:} Bitstrings $\mathbf{x}$, Hamiltonian $\hat{H}$, number of batches $k$, number of iterations $N_{\mathrm{iter}}$, spin and particle number targets
\State \textbf{Output:} Approximate eigenvalue $E^{(k)}$, eigenstate $|\psi^{(k)}\rangle$

\State \textit{\textbf{Initialize carryover sets: $\mathcal{C}_\alpha \gets \emptyset$, $\mathcal{C}_\beta \gets \emptyset$} and cutoff $\epsilon$}

\For{iteration $= 1$ to $N_{\mathrm{iter}}$}
    \If{iteration $= 1$}
        \State Discard bitstrings with incorrect particle number 
    \Else
        \State Modify bitstrings via bit-flip correction using previous $n_{p\sigma}$ to match target symmetry
    \EndIf

    \State Group valid bitstrings into batches: $\mathcal{S}^{(k)}$
    \For{each batch $\mathcal{S}^{(k)}$}
        \State \textit{\textbf{Extend $\mathcal{S}^{(k)}$ with unique carryover bitstrings}: $\mathcal{C}_\alpha$, $\mathcal{C}_\beta\}$}:   $\mathcal{S}^{(k)}\cup(\mathcal{C}_\alpha\otimes\mathcal{C}_\beta)\to \tilde{\mathcal{S}}^{(k)}\}$
        \State Project the Hamiltonian: $\hat{H}_{\tilde{\mathcal{S}}^{(k)}} = \hat{P}_{\tilde{\mathcal{S}}^{(k)}} \hat{H} \hat{P}_{\tilde{\mathcal{S}}^{(k)}}$
        \State Solve eigenvalue equation with spin penalty:
        $\left(\hat{H}_{\tilde{\mathcal{S}}^{(k)}} + \lambda(\hat S^{2}_{\mathcal S^{(k)}} - s(s+1))^2 \right) |\psi^{(k)}\rangle = E^{(k)} |\psi^{(k)}\rangle$
        \State Compute spin-resolved orbital occupancy: 
        $n_{p\sigma} = \langle \psi^{(k)} | \hat{a}^\dagger_{p\sigma} \hat{a}_{p\sigma} | \psi^{(k)} \rangle$
    \EndFor

    \State Select the lowest-energy eigenstate among all batches
    \State \textit{\textbf{Extract high-weight configurations (bitstrings) from the lowest-energy batch for carryover:
    $\mathcal{C}_\alpha, \mathcal{C}_\beta \gets \mathrm{Top determinants with } |\mathrm{coefficient}| > \epsilon
    $}}
    \State Update orbital occupancy for next iteration using averaged $n_{p\sigma}$ across batches

\EndFor

\State \Return $E^{(k)}, |\psi^{(k)}\rangle$
\end{algorithmic}
\end{algorithm}
The bolded and italicized text in Algorithm \ref{alg:co_sqd} highlights the differences compared to the conventional SQD in Algorithm \ref{alg:sqd}.
\subsubsection{Local Unitary Cluster Jastrow (LUCJ) Ansatz}
SQD uses a quantum circuit to sample configurations. In this work, we choose the local-unitary cluster Jastrow (LUCJ) ansatz \cite{LUCJ}. LUCJ is derived from the unitary cluster Jastrow (UCJ) ansatz \cite{UCJ}, which is the unitary variant of the cluster Jastrow (CJ) ansatz~\cite{EN_CJ}.

The UCJ ansatz is a product of $L$ layers of exponentials:
\begin{equation}\label{ucj}
    |\Psi\rangle=\prod_{\mu=1}^L e^{\hat{K}_\mu} e^{i \hat{J}_\mu} e^{-\hat{K}_\mu}\left|\Phi_{0}\right\rangle,
\end{equation}
where $\Phi_{0}$ is a reference state, which here is the Hartree-Fock state. Here, $\hat{K}_\mu$
\begin{equation}
    \hat{K}_\mu=\sum_{p q, \sigma} K_{p q}^\mu \hat{a}_{p \sigma}^{\dagger} \hat{a}_{q \sigma}, \hat{J}_\mu=\sum_{p q, \sigma \tau} J_{p \sigma, q \tau}^\mu \hat{n}_{p \sigma} \hat{n}_{q \tau}.
\end{equation}
is an orbital rotation operator, and each $\hat{J}_\mu$ is a diagonal Coulomb operator. $p,q = 0,...,N_{\mathrm{MO}}-1$, and $\sigma,\tau$ label spin polarizations.

LUCJ introduces a ``local" approximation of the UCJ ansatz~\cite{Motta2021} that allows opposite spin interactions at the same orbital and linear interactions of the same spin from the adjacent orbital:
\begin{equation}
\begin{aligned}
& \sum_{p q} J_{p \alpha, q \beta} \hat{n}_{p \alpha} \hat{n}_{q \beta} \rightarrow \sum_{p \in S} J_{p \alpha, p \beta} \hat{n}_{p \alpha} \hat{n}_{p \beta} \\
& \sum_{p q} J_{p \sigma, q \sigma} \hat{n}_{p \sigma} \hat{n}_{q \sigma} \rightarrow \sum_{p q \in S^{\prime}} J_{p \sigma, q \sigma} \hat{n}_{p \sigma} \hat{n}_{q \sigma}.
\end{aligned}
\end{equation}
Here, $\sigma \in {\alpha, \beta}$; $S,  S^{\prime}$ are sets of qubits that encode the occupation of spin-up and spin-down orbitals. On IBM's heavy-hex processor, for instance, $S = \{4k, k =0,...,\lfloor{\frac{N_{\mathrm{MO}}-1}{4}\rfloor\}}$,$S^{\prime} = \{(p,p+1), p=0,...,N_{\mathrm{MO}}-2\}$. The implementation of $e^{i \hat{J}_\mu}$ is of constant depth ($\mathcal{O}(1)$), since it involves exponentiation of the identity and Pauli Zs. The implementation of $e^{\pm\hat{K}_\mu}$ uses a Bogolyubov circuit~\cite{BogoliubovTransform1,BogoliubovTransform2}, which scales as $\mathcal{O}(N_q)$ in depth, with $N_q$ being the number of qubits.

\subsection{LASSQD}
\subsubsection{RDM-based LASSCF\label{las_rdm}}
The LASSCF implementation in the package mrh~\cite{mrh} allows substitution of the local fragment FCI wave function solver with any approximate wave function solver, provided the corresponding 1- and 2-RDMs are also available to drive the orbital optimization. In conventional LASSCF, FCI is performed for the $K$th fragment in the presence of an external potential defined by the RDMs of the other fragments in the microcycle, and this potential is updated every time FCI is performed. In this work, we lightly modify this implementation: the external potentials are precomputed before the loop over fragments, enabling independent fragment-local CI calculations and thus allowing parallel computation of the fragment wave functions on quantum hardware. Orbital rotations in the full molecular orbital (MO) space are optimized during the macro cycles, except for orbital rotations within fragments, which become nonredundant for approximate wave functions, and are not considered in the LASSCF orbital optimization. 

Specifically, the micro cycles correspond to solving each fragment’s electronic problem using the SQD solver, where the $\hat{U}_{\mathrm{CI}, K}$ are updated while keeping the molecular orbitals fixed. Once all fragment solvers converge, their 1- and 2-RDMs are assembled. The macro cycle then updates the molecular orbitals $ \hat{U}_{\mathrm{orb}}$ using these RDMs to minimize the total energy, analogous to the orbital-optimization step in LASSCF. 
\subsubsection{LASSQD}\label{lassqd}
In this work, we use the SQD algorithm (and its carryover variant), to return fragment 1- and 2-RDMs to the LASSCF solver for orbital optimization. Specifically, a macro-iteration consists of: (i) generating all fragment circuits, (ii) hardware execution and bitstring retrieval, and (iii) classical post-processing with SQD to process hardware results and update orbitals. Repeating this loop over several macro-iterations yields fragment solutions and orbital optimization consistent with the LASSCF framework. To integrate the carryover workflow, we save the MO coefficient matrix for each fragment and perform the standard carryover SQD procedure. At its completion, we obtain both the carryover bitstrings and the MO coefficient matrices for subsequent iterations. At the next iteration, the previously saved MO coefficients are reloaded, and the overlap matrix between the current and prior MO bases is computed for each fragment. The saved $\alpha$ and $\beta$ strings are then permuted according to the optimal overlap mapping and used as input configurations in the next SQD cycle. 


\subsubsection{LASSCI}
To compare the accuracy of LASSQD results when solving for fragment sizes intractable to LASSCF, we further take advantage of the RDM-based LASSCF framework and implement a classical SCI solver, LASSCI, for the fragments. We use the SCI as implemented in PySCF v2.10 (the pyscf.fci.selected\_ci module) 
Specifically, LASSCI constructs a solver object for each fragment in sequence, applies a spin constraint, and then diagonalizes with the Davidson algorithm to obtain the lowest-energy CI eigenstates and coefficients. From the leading root, it constructs spin-resolved 1- and 2-RDMs, as well as the expected spin-squared value for verification. The only difference between classical LASSCI and LASSQD consists in the way in which the SCI space is built.
\subsection{LASSQD-PDFT}
Active-space methods capture the essential static correlation in the multireference systems but inherently miss dynamical correlation, regardless of how accurate the active-space treatment is. Here, we retrieve it through classical post-processing based on densities exclusively, namely the multiconﬁguration pair-density functional theory (MC-PDFT). MC-PDFT is a post-SCF treatment that uses the kinetic energy, electron density, and on-top pair density of a multiconfigurational wave function. MC-PDFT expresses the total energy for a single state as ~\cite{mcpdft,mcpdft2,mcpdft3,laspdft}:
\begin{equation}
\begin{aligned}
E_{\mathrm{MC}-\mathrm{PDFT}}&[\rho, \Pi]=  V_{\mathrm{NN}}+T_{\mathrm{MC}}+\int \rho(\mathbf{r}) \nu_{\mathrm{Ne}}(\mathbf{r}) \mathrm{d} \mathbf{r} \\
& +\frac{1}{2} \iint \frac{\rho\left(\mathbf{r}_1\right) \rho\left(\mathbf{r}_2\right)}{r_{12}} \mathrm{~d} \mathbf{r}_1 \mathrm{~d} \mathbf{r}_2+E_{\mathrm{ot}}[\rho, \Pi],
\end{aligned}
\end{equation}
where $V_{\mathrm{NN}}$ is the nuclear repulsion energy, $T_{\mathrm{MC}}$ the kinetic energy of the multiconfigurational reference wave function, $\langle\Psi^\mathrm{MC}|\hat{T}|\Psi^\mathrm{MC}\rangle$, $\int \rho(\mathbf{r}) \nu_{\mathrm{Ne}}(\mathbf{r}) \mathrm{d} \mathbf{r}$ the nuclear-electron attraction term and $\frac{1}{2} \iint \frac{\rho\left(\mathbf{r}_1\right) \rho\left(\mathbf{r}_2\right)}{r_{12}} \mathrm{~d} \mathbf{r}_1 \mathrm{~d} \mathbf{r}_2$ is the Coulomb repulsion calculated from the wave function density $\rho$.
These four terms represent the classical energy.
The last term, $E_{\mathrm{ot}}[\rho, \Pi]$, called the on-top energy, is the non-classical energy. It is a functional of $\rho$ and the on-top pair density $\Pi$ of the reference wave function. In this work, the multireference wave function is $\Psi_{\mathrm{LASSQD}}$. 

In our implementation, each fragment's 1- and 2-RDMs are stored during the LASSQD calculation for post-LASSQD analysis. The $\rho$ and $\Pi$ term depend on the 1- and 2-RDMs of the whole molecule, which are obtained by assembling the 1- and 2-RDMs of the fragments in a straightforward way.

\subsection{Computational details}
\begin{table*}[t]
\centering
\renewcommand{\arraystretch}{1.3}
\setlength{\tabcolsep}{7pt}
\caption{The molecular systems used in this work, with their geometry, basis set, fragmentation and number of shots used for hardware calculations.}
\begin{tabular}{ccccc}
\hline\hline
\textbf{System}  & \textbf{Basis set} & \textbf{Geometry} & \textbf{Fragment} & \textbf{Shots} \\
\hline
[Fe(H$_2$O)$_4$]$_2$bpym$^{4+}$  & 6-31G & Ref. \cite{vLASSCF} & ((6e,5o),(6e,5o)) & \scientific{3}{4} \\
--- &  6-31G & Ref. \cite{vLASSCF} & ((6e,10o),(6e,10o)) & \scientific{8}{4} \\
$[\mathrm{Fe}^{\mathrm{III}}\mathrm{Fe}^{\mathrm{III}}\mathrm{Fe}^{\mathrm{II}}(\mu_3\text{-}\mathrm{O)\text{-}(HCOO)}_6]$
 & cc-pVDZ (Fe), 6-31G & Ref. \cite{doi:10.1021/acscatal.8b04813} & ((6e,10o),(6e,10o),(6e,10o)) &\scientific{9}{4}  \\
iron-porphyrin & 6-31G & Ref. \cite{fciqmc_scf1} & ((6e,5o),(26e,23o)) & \scientific{8}{4} \\
\hline
\end{tabular}
\label{tab:comp_detail}
\end{table*}
Calculations performed on the classical computers are carried out with PySCF~\cite{Pyscf, PySCF2} version 2.10 and the mrh package~\cite{mrh}. LASSQD calculations with samples from QASM (classical noiseless simulator of quantum circuits) and quantum hardware are carried out with Qiskit~\cite{qiskit2024}, SQD addon software~\cite{qiskit-addon-sqd}, while LUCJ circuits are constructed with ffsim~\cite{ffsim} software library, and the code used in this work is available at LASSQD repository~\cite{lassqd_code}. Dynamical decoupling~\cite{DD1, dd2,DD3} and gate twirling~\cite{PT} are used for all hardware calculations for error mitigation. The geometry, basis set, fragmentation and shot usage are listed in Table \ref{tab:comp_detail}. Each system's spin multiplicity and SQD settings are described in the corresponding results section.

All circuits used in the calculation are truncated LUCJ circuits in the form of:
\begin{equation}
    |\Psi\rangle= e^{\hat{K}_2}e^{-\hat{K}_1} e^{i \hat{J}_1} e^{\hat{K}_1}\left|\mathbf{x}_{\mathrm{RHF}}\right\rangle,
\end{equation}which has the same depth of Eq. \ref{ucj} when $L=1$. LUCJ circuit parameters were fully optimized at each macro iteration with the linear method~\cite{ToulouseUmrigar,MOTTA201562,LMLUCJ} in Section \ref{SQD_is_FCI}, \ref{SQD_less_FCI}, and the results in Section \ref{IP} are obtained directly from the $t_1$ and $t_2$ amplitudes from CCSD calculations, as described in Ref. \cite{SQD}. The results section contains information on specific quantum hardware used and qubit usage. In this work, we assess the accuracy of LASSQD based on total electronic energies  (in \textit{E\textsubscript{h}}) rather than relative energy differences (in \textit{mE\textsubscript{h}} or kcal/mol). This choice allows a more direct evaluation of the variational quality of LASSQD and LASSCI compared to LASSCF, and all the methods mentioned above obey the variational principle.

\section{Results and discussions}
\subsection{SQD equivalent to FCI in small LAS fragments\label{SQD_is_FCI}}
To understand the accuracy of our method, we compare LASSQD results against classical LASSCF, for which FCI is the fragment solver and therefore is exact in the fragments. For sufficiently small fragments, SQD reproduces FCI exactly, and therefore we expect LASSQD to reproduce LASSCF exactly. We perform LASSQD hardware experiments (Figure 1. b) for [Fe(H$_2$O)$_4$]$_2$bpym$^{4+}$ (Figure 1. a). We use 15 batches ($K=15$), 6 iterations ($N_\mathrm{iter}=6$), samples per batch to 10 ($d$ =10). Each fragment is set up with a different number of $\alpha$ (spin-up) and $\beta$ (spin-down) electrons; fragment 1 contains 4 $\alpha$ electrons and 2 $\beta$ electrons, while fragment 2 has 2 $\alpha$ electrons and 4 $\beta$ electrons. Since each fragment is of size (6e,5o), the bitstrings sampled from the circuits can cover full set of configurations, and thus for SQD to diagonalize the full Hilbert space Hamiltonian.  Our results in Figure 1. c and Table 1 show that LASSQD achieves near-exact agreement with LASSCF, with an energy difference of just $5.3 \cdot 10^{-7}$ \textit{E\textsubscript{h}}, demonstrating the high accuracy that can be obtained with the method for small active space sizes. We further compare the PDFT energies from the LASSCF wave function and LASSQD-PDFT. As Table \ref{fefe_table1} shows, the two values are again in agreement to  $1\cdot 10^{-6}$. 
\begin{table}[h!]
\renewcommand{\arraystretch}{1.3}
\setlength{\tabcolsep}{7.5pt}
\caption{\label{fefe_table1}Comparison of the absolute energy of [Fe(H$_2$O)$_4$]$_2$bpym$^{4+}$ intermediate spin state with various methods.}
    \begin{tabular}{lc}
    \hline\hline
    \multicolumn{1}{c}{\bf{Method}}&
    \multicolumn{1}{c}{\bf{Energy (\textit{E\textsubscript{h}})}}\\
    \hline
    LASSQD ((5o,6e),(5o,6e)) & -3655.496377  \\
    LASSCF ((5o,6e),(5o,6e))& -3655.496377  \\
    LASSQD-tPBE& -3662.471086 \\
    LAS-tPBE& -3662.471085 \\
    \hline
    \end{tabular}
    
\end{table}

\begin{figure*}[]
    \centering
\includegraphics[width=0.9\linewidth]{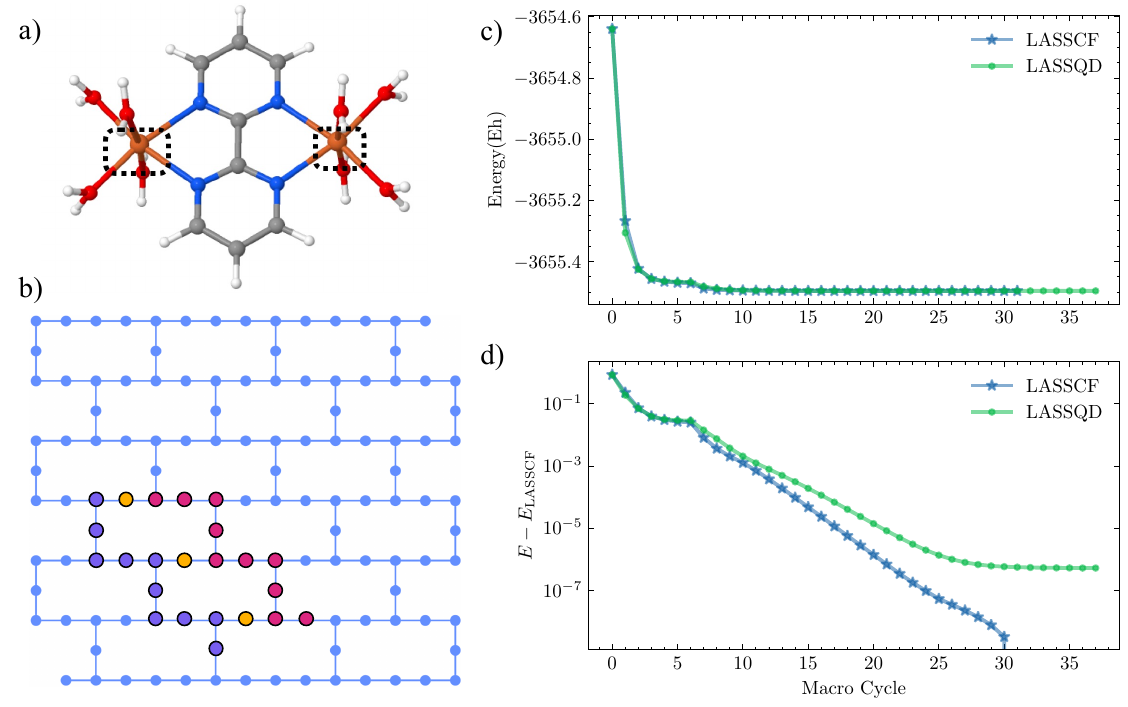}
    \caption{\textbf{a)}: Bimetallic compound [Fe(H$_2$O)$_4$]$_2$bpym $^{4+}$, fragmentation illustrated in dotted box. Each fragment contains 5 3$d$ orbitals and 6 electrons. The first fragment has 4 spin-up electrons and 2 spin-down electrons, while the second fragment has 2 spin-up electrons and 4 spin-down electrons. \textbf{b)}: Qubits used on \textit{ibm\_sherbrooke} are indicated by red (purple) circles, which correspond to $\alpha$ ($\beta$) spin orbitals. Orange encodes auxiliary qubits that mediate the density-density interactions among orbitals of opposite spin in the LUCJ ansatz. \textbf{c)}: absolute energy plots of LASSCF and LASSQD energies from hardware across consecutive macro cycles. \textbf{d)}: energy difference from each macro cycles of classical LASSCF and LASSQD hardware calculations in log scale with the converged LASSCF energy ($E_\mathrm{LASSCF}$ in the legend) as the reference. }
    \label{fig:sec1}
\end{figure*}

In the standard SQD solver, the Davidson algorithm searches for only a single root with a default residual norm tolerance (defaulted at 10$^{-9}$), which can cause convergence to a low-lying excited state rather than the true ground state. This is particularly problematic in our case of the intermediate spin configuration, where the triplet configuration may have nearly degenerate eigenstates with each other. The SQD solver can converge when the residual vector norm becomes sufficiently small, even if the variational subspace does not yet contain the true ground state. To address this, we make improvements to the original SQD solver by allowing multiple root searches in a single calculation and lowering the tolerance to $10^{-16}$. This allows us to successfully find the exact ground state in the fragments and yield the same energy as LASSCF as reported in 
Figure \ref{fig:sec1} and Table \ref{fefe_table1}. See SI Section II for more details.

\subsection{SQD approximates FCI in LAS fragments\label{SQD_less_FCI}}
The full Hilbert space grows exponentially with system size, making exact diagonalization classically intractable for most systems of interest. The core of SQD lies in the selection of a compact, physically relevant subspace of determinants in which the Hamiltonian is diagonalized, following the philosophy of SCI methods. By restricting the diagonalization to a judiciously chosen subset of determinants, SQD may provide an efficient approximation to FCI that could be well-suited for a fragment solver in LASSCF. To demonstrate that LASSQD provides an efficient approximation to FCI in the fragments while enabling meaningful orbital optimizations, we enlarge the fragment active space for [Fe(H$_2$O)$_4$]$_2$bpym$^{4+}$ to 10 orbitals, including 5 3$d$ orbitals and 5 4$d$ orbitals, and 6 electrons in each fragment in a triplet configuration (see Figure \ref{fig:sec1} a)). This specific fragment AS size is still classically tractable, which allows a comparison between LASSQD performance to LASSCF when the LASSQD fragment solution is no longer equal to FCI. 
\subsubsection{Conventional SQD for LASSQD}

\begin{figure}[h]
\includegraphics[width=1\linewidth]{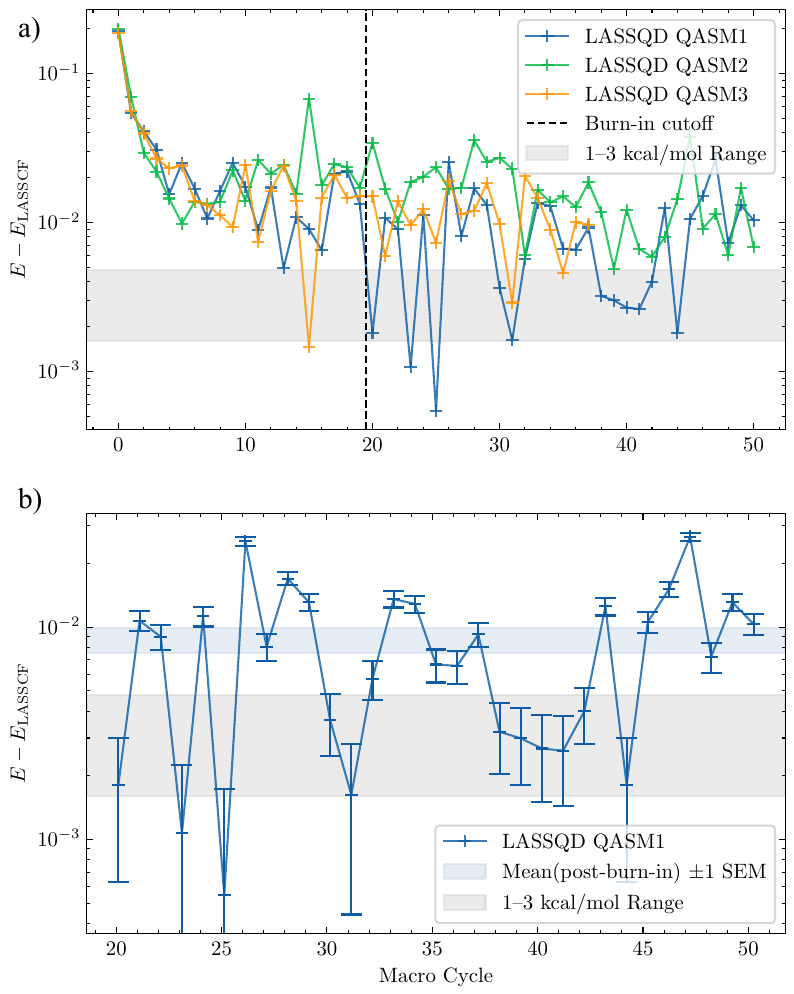}
    \caption{\textbf{a)} Energy difference with error bars between LASSQD QASM simulations and converged LASSCF energy ($E_\mathrm{LASSCF}$ in the legend) as the reference. 1-3 kcal/mol deviations from LASSCF energy is plotted as an acceptable range of error. The dashed vertical line indicates the cutoff for statistical analysis. \textbf{b)}: Energy difference with error bars between LASSQD QASM1 simulation and LASSCF in log scale. The light blue box shows the ensemble's mean with 1 standard error after the burn-in cutoff.  }
    \label{fig:sec2_noco}
\end{figure}

Here we first present results from the conventional SQD implementation of LASSQD for [Fe(H$_2$O)$_4$]$_2$bpym$^{4+}$ with three different QASM simulations (LASSQD QASM 1, QASM 2, QASM 3). We use 15 batches ($K=15$), 6 iterations ($N_\mathrm{iter}=6$), samples per batch to 170 ($d$ =170), which is around 80\% of the maximum $\alpha$($\beta$) configurations. QASM 1 uses \scientific{1}{5} shots and fully optimized LUCJ circuit parameters with the linear method, with $n_{root} = 10$ and tolerance set to \scientific{1}{-16} for the Davidson algorithm. QASM 2 has \scientific{8}{4} shots and a maximum optimization iteration to 20, with $n_{root} = 3$ and tolerance set to \scientific{1}{-16}. QASM 3 is carried out with \scientific{8}{4} shots, maximum optimization iteration of 40, $n_{root} = 5$, and tolerance set to  \scientific{1}{-16}. The energy differences to reference LASSCF (converged LASSCF energy) from the above three simulations are plotted in Figure \ref{fig:sec2_noco} top panels.
As Figure \ref{fig:sec2_noco} top panel shows, all three simulations exhibit stochastic behavior, arising from the new sampling at every macro iteration. Although LASSQD is not a purely stochastic algorithm, as the orbital optimization followed by the CI update remains deterministic, the conventional SQD implementation produces characteristic “zig-zag” oscillations, with energy errors on the order of $10^{-2}$ to $10^{-3}$ Eh. 
To evaluate whether these oscillations fall within an acceptable energy range, we perform a statistical analysis of the LASSQD QASM 1 simulations. After visually inspecting the trajectories, we discard the first 20 data points (denoted as “Burn-in” in the legend), which likely correspond to the transition phase before the algorithm approaches a reasonable orbital minimum. We then calculate the mean and the standard error of the mean (SEM) for the remaining ensemble and plot the energy differences with error bars in the lower panels of Figure \ref{fig:sec2_noco}.
While some of the individual data points from the LASSQD QASM1 simulations enter the 1-3 kcal/mol range relative to LASSCF, the post-burn-in mean lies more than 3 kcal/mol above the reference, indicating that the conventional SQD approach requires refinement to be practical in an iterative SCF framework. Moreover, the consistency of oscillations across three QASM simulations with different parameters suggests that the cause stems from a fundamental aspect of the SQD implementation, rather than from parameter-specific numerical noise. This motivates the development of carryover SQD, discussed in the following section.

\subsubsection{Carryover SQD for LASSQD}\label{carryoverlassqd}
The underlying issues identified in the previous sections can be summarized as follows:

1) At each macro cycle, we sample from a new circuit without retaining information about the bitstrings obtained in earlier cycles. As a result, diagonalization in this newly sampled subspace does not necessarily yield a lower energy solution, even after orbital optimization.

2) A large number of shots is required to capture all important configurations that contribute significantly to the energy expression. 

While both challenges are intrinsic to SQD, we focus here on addressing the first. To this end, we use a retention step, in which important bitstrings are saved and carried over from one iteration to the next. This ensures that the diagonalization consistently incorporates the most relevant contributions.
\begin{figure*}
    \centering
\includegraphics[width=1\linewidth]{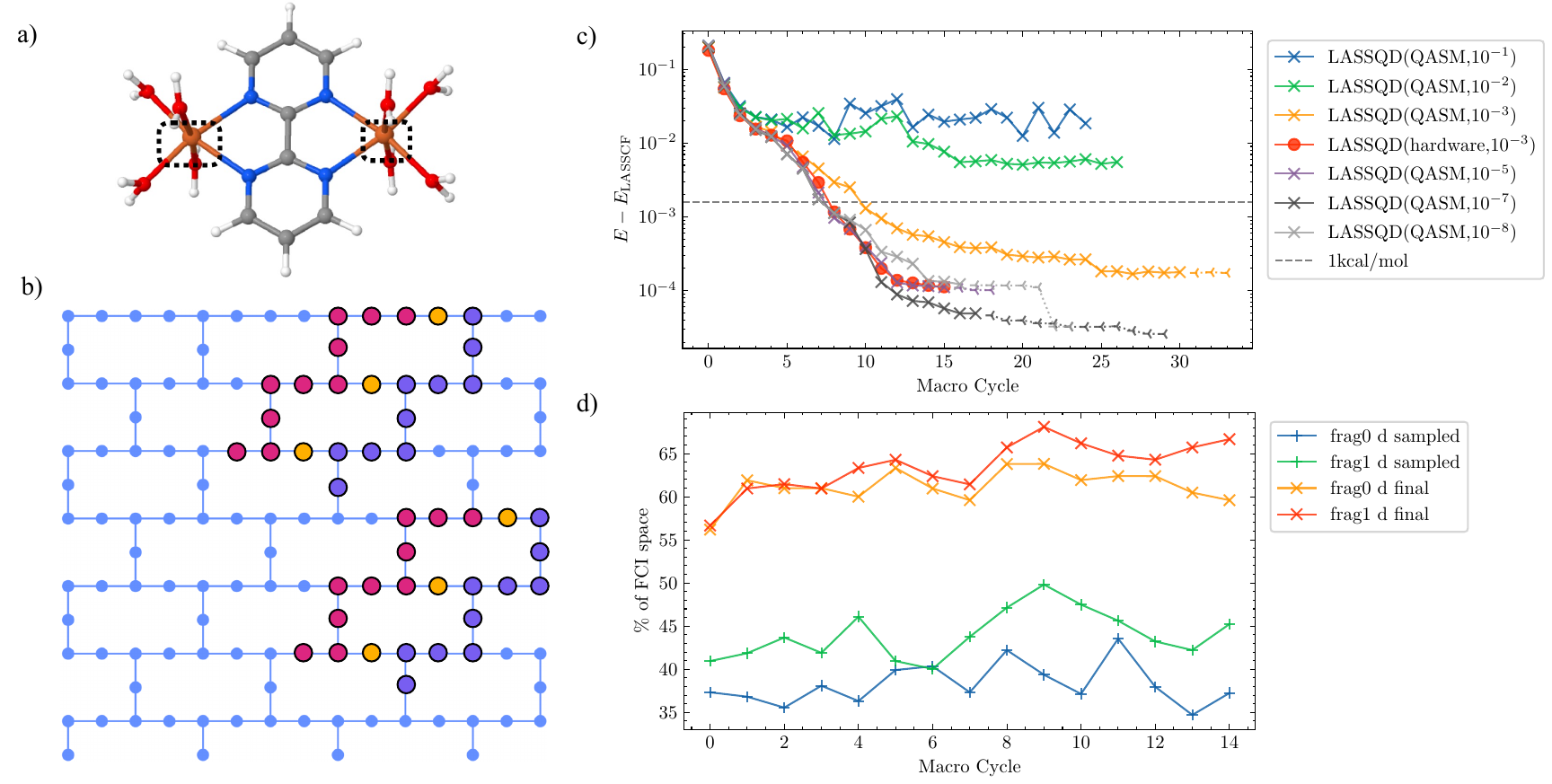}
\caption{\textbf{a)}: Bimetallic compound [Fe(H$_2$O)$_4$]$_2$bpym$^{4+}$, fragmentation illustrated in dotted box. Each fragment contains 10 orbitals (5 3$d$ orbitals and 5 4$d$ orbitals) and 6 electrons. The first fragment has 4 spin-up electrons and 2 spin-down electrons, while the second fragment has 2 spin-up electrons and 4 spin-down electrons. \textbf{b)}: Qubits used on \textit{ibm\_torino} are indicated by red (purple) circles, which correspond to $\alpha$ ($\beta$) spin orbitals. Orange encodes auxiliary qubits that mediate the density-density interactions among orbitals of opposite spin in the LUCJ ansatz. \textbf{c)}: LASSQD energy difference in log scale to the reference LASSCF energy. The dashed grey line is 1 kcal/mol deviation from LASSCF. The legend indicates the backend type (QASM or hardware), and the carryover threshold in powers of 10. \textbf{d)}: The maximum subspace dimension as the percentage of FCI space plotted across each macro cycle for LASSQD hardware calculation. “d sampled” denotes the raw SQD dimension and “d final” denotes the final subspace after carryover}
\label{fig:sec2_combof3}
\end{figure*}

With carryover SQD, where key bitstrings are retained from macro cycle $n$ to $n+1$, the algorithm exhibits improved behavior. Using the qubit configuration in Figure \ref{fig:sec2_combof3} \textbf{b)}, hardware calculations (Figure \ref{fig:sec2_combof3} \textbf{c)}, red line) reach sub–1 kcal/mol accuracy after nine macro cycles and converge smoothly with orbital optimization. The setup employs $K=15$ batches, $N_\mathrm{iter}=6$, and $d=170$ samples per batch, covering ~80\% of the maximum 
$\alpha(\beta)$ configurations. Figure \ref{fig:sec2_combof3} \textbf{d)} shows the maximum SQD dimension per macro cycle as a fraction of the FCI space, with “d sampled” denoting the raw SQD dimension and “d final” denoting the final subspace after carryover.

\begin{align}
    d~ \text{sampled} &= \max_k|\mathcal{S}^{(k)}| \\
    d &= |\tilde{\mathcal{S}}^{(k_{\textrm{final}})}|
\end{align}
where, in the final SQD cycle of a given orbital-optimization cycle,
\begin{equation}
    k_{\textrm{final}} \equiv \textrm{arg}\min_k E^{(k)}
\end{equation}

The overall subspace (sub-70\% of the FCI space), shows that LASSQD is efficient in approximating FCI for bimetallic systems, and it provides an approach for high-accuracy calculations with orbital optimization when FCI is intractable. However, we still need to investigate three aspects of the hardware calculations: 1) whether the LASSQD orbital optimization can reach a similar minimum to LASSCF, 2) if similar results can be obtained for other systems, and 3) whether the utilization of the subspace dimension of SQD is efficient. To further answer these questions, we carry out a series of LASSQD QASM simulations (Figure \ref{fig:sec2_combof3} \textbf{c)}) to perform localized active space configuration interaction (LASCI) calculations with orbitals from the aforementioned LASSQD simulations (Table \ref{fefe_table2}), LASSQD before and after carryover simulations for [Fe$^{\mathrm{III}}$Fe$^{\mathrm{III}}$Fe$^{\mathrm{II}}$($\mu$$_3$-O)-(HCOO)$_6$], and LASSCI calculations as comparison in Section \ref{sec:sqdvsci} Table \ref{fefe_table3}.  

From Figure \ref{fig:sec2_combof3} \textbf{c)}, we first observe that when the carryover cutoff ($\epsilon$) is small (e.g., \scientific{1}{-1}, \scientific{1}{-2}), the absolute energies appear stochastic, which is similar to when there are no bitstrings retained from one macro cycle to the next (Figure \ref{fig:sec2_noco}). In other words, due to sampling from a new circuit at every macro iteration, when too few bitstrings are retained, the energy trend shows stochasticity. We also find that the hardware LASSQD calculation and the rest of the simulations reach convergence, which we define using the criterion $\Delta E < 1\cdot10^{-5}$ \textit{E\textsubscript{h}} for two consecutive energy values, where $\Delta E$ is the difference in LASSQD energy from macro $n$ to $n+1$ cycle. This criterion serves as a meaningful stopping point for comparison of absolute energies and SQD dimensions in the following analysis. QASM energy values beyond convergence are plotted with dotted lines in Figure \ref{fig:sec2_combof3} \textbf{c)}. 
\begin{table}[h!]
\renewcommand{\arraystretch}{1.3}
\setlength{\tabcolsep}{7.5pt}
\caption{\label{fefe_table2} Absolute energies for LASSQD QASM simulations in Figure \ref{fig:sec2_combof3} d) for system [Fe(H$_2$O)$_4$]$_2$bpym$^{4+}$, and subsequent LASCI calculations with LASSQD orbitals. $n$ represents the macro cycle number. }
    \begin{tabular}{lc}
    \hline\hline
    \multicolumn{1}{c}{\bf{Method}}&
    \multicolumn{1}{c}{\bf{Energy (\textit{E\textsubscript{h}}) }}\\
    \hline
    LASSQD$_{1}$ ($\epsilon=$\scientific{1}{-3}) & -3655.610803 ($n=33$)
     \\
    LASSQD$_{2}$ ($\epsilon=$\scientific{1}{-5})& -3655.610875 ($n=18$)\\
    LASSQD$_{3}$ ($\epsilon=$\scientific{1}{-7}) & -3655.610951 ($n=29$)  \\
    LASSQD$_{4}$ ($\epsilon=$\scientific{1}{-8})&
    -3655.610944 ($n=23$)\\
    LASCI(LASSQD$_{1}$\textsuperscript{\emph{a}}) &-3655.610977 \\
    LASCI(LASSQD$_{2}$\textsuperscript{\emph{a}})& -3655.610977 \\
    LASCI(LASSQD$_{3}$\textsuperscript{\emph{a}})& -3655.610977 \\
    LASCI(LASSQD$_{4}$\textsuperscript{\emph{a}})& -3655.610977 \\
    LASSCF & -3655.610977\\
    \hline
    \end{tabular}
    
    \textsuperscript{\emph{a}} LASSQD orbitals
\end{table}

We further take the LASSQD orbitals at the end of the calculations and use them as guess orbitals to perform subsequent LASCI calculations. The LASCI energy difference from LASSQD should only include the difference between FCI and the SQD approximation. In other words, by examining how close LASCI energies are to LASSCF, we can understand if the LASSQD orbitals are in a similar minimum as LASSCF. From Table \ref{fefe_table2}, we see a near-exact agreement (to \scientific{1}{-7} \textit{E\textsubscript{h}}) among the LASCI calculations and the LASSCF calculation. It means that the various LASSQD calculations, though differing in final energies, carryover cutoffs, and therefore the SQD dimensions, all converge to the same set of orbitals and are in agreement with LASSCF. It affirms that LASSQD orbitals, despite SQD approximating FCI in the fragments, can optimize to the same set of orbitals as LASSCF. This data supports the point that the SQD approximation to FCI still allows for meaningful orbital optimization. 

\begin{figure}
\centering\includegraphics[width=\linewidth]{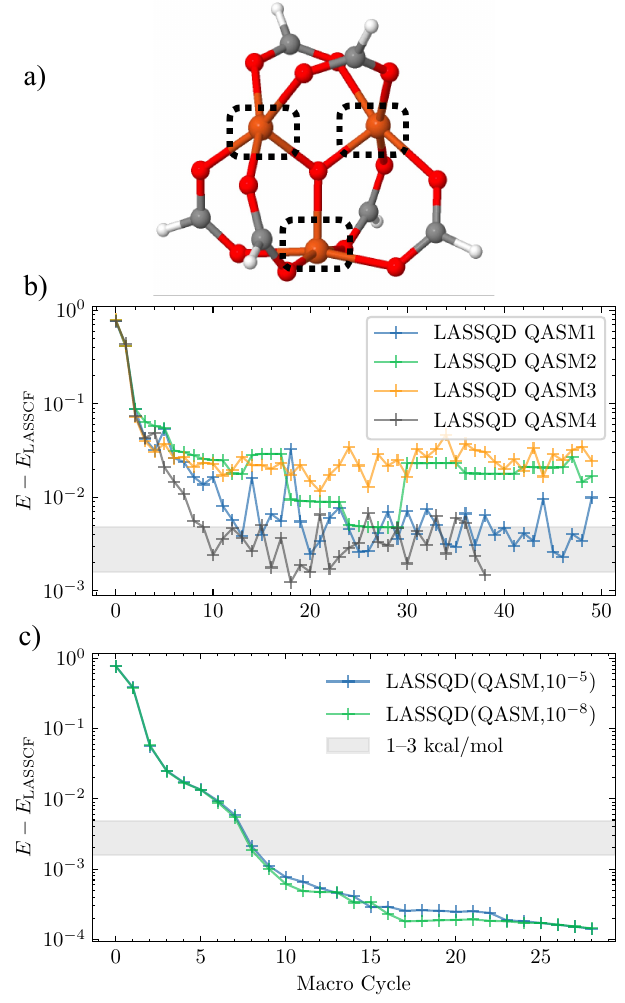}
    \caption{\textbf{a)}: Triiron oxo-centered complex [Fe$^{\mathrm{III}}$Fe$^{\mathrm{III}}$Fe$^{\mathrm{II}}$($\mu$$_3$-O)-(HCOO)$_6$], with dashed box illustrating fragmentation. Each fragment has 10 orbitals, and (5$\alpha$, 1$\beta$),(4$\alpha$, 1$\beta$),(4$\alpha$, 1$\beta$) electrons respectively. \textbf{b)}: LASSQD QASM simulations without carryover, shaded grey area indicates 1-3 kcal/mol accuracy with respect to LASSCF. \textbf{c)}: LASSQD QASM simulations with carryover, $\epsilon=$\scientific{1}{-5}, \scientific{1}{-8}. }
    \label{fig:FE3O}
\end{figure}

To confirm that this is not an isolated case with the [Fe(H$_2$O)$_4$]$_2$bpym$^{4+}$ compound, we tested LASSQD before and after carryover on a more challenging case with three strongly correlated fragments. Here we present results of the LASSQD QASM simulation of the triiron oxo-centered complex [Fe$^{\mathrm{III}}$Fe$^{\mathrm{III}}$Fe$^{\mathrm{II}}$($\mu$$_3$-O)-(HCOO)$_6$] ((10o,6e),(10o,5e),(10o,5e)) (Figure \ref{fig:FE3O}, \textbf{a)}). The Fe$_3$O system has three iron centers, which are the three fragments in our calculation. Each fragment contains 5 3$d$ orbitals and 5 4$d$ orbitals, with one iron center having 6 electrons and the other two having 5 electrons. All iron centers are paragnetic, and have unbalanced $\alpha$ and $\beta$ electrons ((5$\alpha$, 1$\beta$),(4$\alpha$, 1$\beta$),(4$\alpha$, 1$\beta$) electrons respectively). We carry out four QASM LASSQD simulations with different settings and parameters as follows: QASM 1 simulation uses \scientific{5}{6} $n_{roots}=10$, QASM 2 uses \scientific{8}{4} shots and $n_{roots}=10$, QASM 3 is carried out with 
\scientific{4}{4} shots and $n_{roots}=6$, and QASM 4 uses 1 million shots and $n_{roots}=10$. All Davidson's tolerances are set to \scientific{1}{-16} except for QASM 3 (tolerance set to \scientific{1}{-14}).
As shown in four QASM simulations \textbf{b)} of Figure \ref{fig:FE3O}, the conventional implementation of SQD exhibits strong oscillations in macro-iteration energies, with errors remaining above 1 kcal/mol relative to LASSCF. In contrast, when using carryover LASSQD, the stochastic fluctuations are substantially reduced: the method surpasses 1 kcal/mol accuracy within 10 macro cycles, and the overall energy profile converges smoothly. With the carryover strategy, LASSQD consistently achieves sub–1 kcal/mol deviations from LASSCF within only a few macro cycles, while simultaneously optimizing orbitals and mitigating sampling noise.
 
\subsubsection{SQD and SCI as fragment solvers}\label{sec:sqdvsci}
Next, we compare the SQD dimensions from LASSQD QASM simulations in Figure \ref{fig:sec2_combof3} to LASSCI calculations to see if SQD generates compact wave functions that are competitive to a classical variant of SCI. We carry out independent LASSQD QASM simulations and LASSCI with various cutoffs, as shown in Appendix A Table \ref{fefe_table3}. Based on results from Figure \ref{fig:sec2_combof3} \textbf{d)}, we use  $\Delta E < 1\cdot10^{-5}$ for two consecutive energy values as the energy and dimension termination points for LASSQD simulations. The subspace size $d$ is taken as the maximum size used in both algorithms.
\begin{figure}[h!]
    \centering
    \includegraphics[width=1\linewidth]{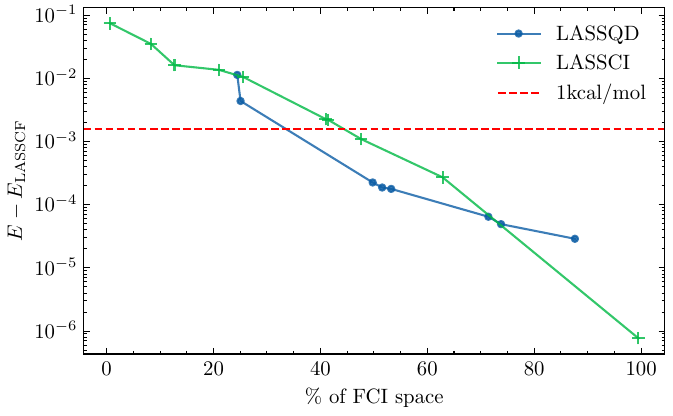}
    \caption{Independent QASM LASSQD runs and classical LASSCI calculations with varying cutoffs. y-axis shows the absolute energies of LASSQD and LASSCI difference from LASSCF while x-axis indicates the percentage of max FCI space used for that energy value. }
    \label{fig:sec2_p4}
\end{figure}

As Figure \ref{fig:sec2_p4} shows, the performance of LASSQD is comparable to LASSCI and within 1 kcal/mol agreement to LASSCF with about $50\%$ of the FCI space. LASSQD simulations also generate lower energies compared to LASSCI when the subspace dimension $d$ is between $~25\% - 70\%$. The energy values and subspace usage data can be found in Appendix \ref{FeFe_dcomp}. This illustrates that LASSQD is competitive against LASSCI and that with carryover, SQD is able to retain useful bitstrings efficiently. This feature is especially important when the FCI space becomes intractable and we have to rely on the sampling and retaining of the SQD algorithm for a useful answer.

\subsection{LASSQD solves beyond classical LASSCF fragment sizes\label{IP}}
From the previous results, we have established encouraging examples where SQD with carryover can approximate FCI efficiently while allowing meaningful optimization.  
We next demonstrate that LASSQD can address systems beyond the reach of LASSCF, yielding results comparable to LASSCI in regimes where LASSCF can no longer be employed.

\begin{figure}
   \centering
    \includegraphics[width=0.85\linewidth]{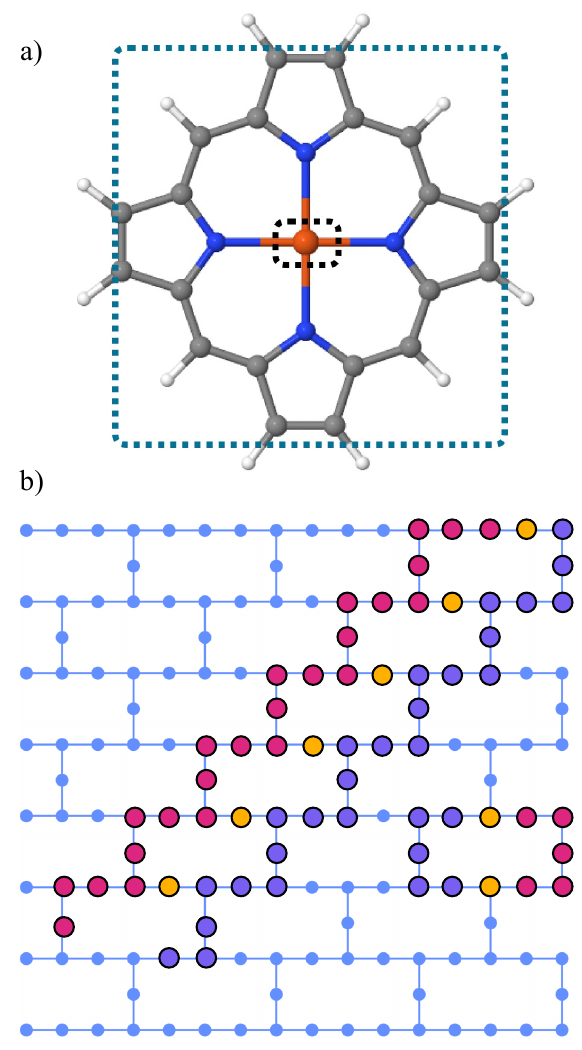}
    \caption{\textbf{a)}: iron-porphryin, fragmentation illustrated in dotted box.  The iron fragment has 6 electrons and 5 $3d$ orbitals (6e,5o), while the second fragment has 26 electrons in 24 $\pi$ orbitals (26e,24o). \textbf{b)}: Qubits used on \textit{ibm\_fez} are indicated by red (purple) circles, which correspond to $\alpha$ ($\beta$) spin orbitals. Orange encodes auxiliary qubits that mediate the density-density interactions among orbitals of opposite spin in the LUCJ ansatz.}
    \label{fig:ip_1}
\end{figure}

Understanding the spin-state energetics of metalloporphyrins with near-degenerate spin states remains a major challenge for computational and theoretical methods. A prominent example is Iron(II) porphyrin (FeP, C$_{20}$H$_{14}$N$_4$Fe). Density functional theory calculations~\cite{IP_DFT1,ip_dft2,ip_dft3}, along with some experimental studies on porphyrin derivatives~\cite{ip_exp1,ip_exp2,ip_exp3,ip_exp4}, predict a triplet ground state. However, the combination of the extensive $\pi$-electron system of the porphyrin and the strongly correlated 3$d$ electrons of the iron center demands prohibitively large active spaces for multireference wave function methods. Limited active-space treatments have incorrectly assigned the quintet as the ground state~\cite{ip_mr_s1,ip_mr_s2,ip_mr_s3,ip_mr_s4}, whereas more recent studies show that expanding the active space and/or including dynamic correlation stabilizes the experimentally observed triplet~\cite{ip_mr_t1,ip_mr_t2,ip_mr_t3}. Our preliminary demonstration aims to test the limit of current hardware capabilities and focuses on computing the spin gap between the lowest spin ($S=0$) and highest spin ($S=2$). This spin gap value will not be compared to experimental values. Instead, we compare the LASSQD predictions with LASSCI.

We use the same total active space (32e, 29o) as in Ref.\cite{fciqmc_scf_is}, and fragment it into the iron atom (6e,5o) and porphyrin ring (26e,24o). We  perform independent LASSCI and LASSQD hardware calculations with different SCI thresholds and carryover cutoffs for both high-spin state and low-spin state. We then carry out energy-variance extrapolation~\cite{ev_extrap} (detailed in SI Section III) for LASSCI and LASSQD. The results are reported in Table \ref{ironp_gap}.

\begin{table}[h!]
\renewcommand{\arraystretch}{1.3}
\setlength{\tabcolsep}{7.5pt}
\caption{\label{ironp_gap} Quintet-singlet gap energy comparisons for iron-porphryin with LASSCI and LASSQD ((6e,5o),(26e,24o)) energy-variance extrapolations.}
    \begin{tabular}{ll}
    \hline\hline
    \multicolumn{1}{c}{\bf{Method}}&
    \multicolumn{1}{l}{\bf{Energy (\textit{E\textsubscript{h}}) }}\\
    \hline

    LASSCI energy-var. extrap. & 0.153064 ($\pm$0.017056)\\
    LASSQD energy-var. extrap. & 0.135391 ($\pm$0.064818)\\

    \hline
    \end{tabular}
    
\end{table}
First, both methods consistently predict the lower energy state to be the high-spin quintet, instead of the singlet low-spin configuration. Second, the two spin-gap predictions are agreeable within their respective uncertainties. The relatively large uncertainties (17 mHa for LASSCI and 65 mHa for LASSQD), however, needs to be addressed. 

In both cases, the uncertainties are calculated as $\sqrt{\sigma_{hs}^2+\sigma_{ls}^2}$, which rely on the qualities of the energy-variance extrapolations for each spin state. For classical LASSCI calculations, obtaining a better linear extrapolation with smaller uncertainties requires running calculations with increasingly large subspaces using the classical Davidson solver, which significantly increases runtime and post-processing complexity. In this work, we collected 12 independent LASSCI HS calculations and 11 independent LASSCI LS calculations with varying cutoffs and subspace sizes, which is sufficient to observe the linear trend, although with additional points the quality of the extrapolations will improve, and result in a lower uncertainty value and a more precise numerical prediction. 

For LASSQD calculations, the hybrid workflow faces practical constraints. Under the limited quantum resource and ecosystem limitations, we performed 5 independent LASSQD calculations for each spin state (10 total), varying both the carryover thresholds and the subspace sizes. While increasing the number of LASSQD data points would certainly improve the extrapolation and narrow the uncertainty, the present results already demonstrate that today’s quantum hardware can execute sufficiently deep circuits and return signals stable enough for reliable classical post-processing, in agreement with classical alternatives. With improved hardware quality, better code partitioning to leverage more HPC resources, and algorithmic improvement, we are confident that LASSQD will deliver increasingly accurate and precise energetic predictions moving forward.

\section{Conclusion}
We introduced LASSQD, a hybrid quantum–classical framework that integrates SQD solvers into LASSCF fragments. This approach represents the first demonstration of using approximate solvers within fragment-based multireference methods, enabling treatment of beyond-FCI–sized fragments while retaining meaningful orbital optimizations. A key innovation is adapting the carryover strategy to the SCF formulation, which mitigates sampling stochasticity, leading to smoother convergence and stable orbital optimization. Applications to bimetallic and trimetallic complexes show that LASSQD efficiently approximates FCI, achieves agreement within 1 kcal/mol of LASSCF using only a fraction of the Hilbert space, and produces results competitive with LASSCI for fragment sizes inaccessible to LASSCF.

Looking ahead, improvements to the SQD solver, such as generalized eigenvalue strategies for larger subspaces or semi-stochastic approaches that leverage sampling probabilities, will further increase scalability. In parallel, extending LASSQD to recover inter-fragment correlations beyond the mean-field level will be essential for treating strongly coupled systems. Together, these developments position LASSQD as a powerful and improvable pathway for applying near-term quantum algorithms to large, strongly correlated systems.

\begin{acknowledgments}
The authors thank Valay Agarawal for their helpful discussions and feedback on this work. The authors acknowledge useful conversations and feedback within the Materials Science Quantum Working Group ~\cite{alexeev2024quantum}.
This work has been supported by the IBM-University of Chicago Quantum Collaboration, under agreement number MAS000364, with access to the fleet of IBM Quantum computers. This work was supported by the U.S. Department of Energy, Office of Science, National Quantum Information Science Research Centers and partially by NSF QuBBE Quantum Leap Challenge Institute (NSF OMA-2121044). The authors thank the Research Computing Center at the University of Chicago for the computing resources.

\end{acknowledgments}
\appendix
\section{\label{FeFe_dcomp}}
\begin{table}[h!]
\renewcommand{\arraystretch}{1.3}
\setlength{\tabcolsep}{7.5pt}
\caption{\label{fefe_table3} Absolute energies for LASSQD QASM simulations with various carryover thresholds and LASSCI reported in Figure \ref{fig:sec2_p4}, and their corresponding subspace sizes as a percentage of the FCI space.}
    \begin{tabular}{lcc}
    \hline\hline
    \multicolumn{1}{c}{\bf{Method}}&
    \multicolumn{1}{c}{\bf{Energy (\textit{E\textsubscript{h}})}}&
    \multicolumn{1}{c}{\bf{Max $d$}}
    \\
    \hline
     LASSQD$_{1}$ ($\epsilon=$ \scientific{1}{-1}) & -3655.59963374 & 24.44\%\\
     LASSQD$_{2}$ ($\epsilon=$\scientific{1}{-2}) & -3655.60659447 & 25.06\% \\
    LASSQD$_{3}$ ($\epsilon=$ \scientific{1}{-9})& 3655.61075183&49.48\% \\
    LASSQD$_{4}$ ($\epsilon=$\scientific{1}{-3})& -3655.61078968&51.56\% \\
    LASSQD$_{6}$ ($\epsilon=$\scientific{1}{-3}) & -3655.61079882 & 53.24\% \\
    LASSQD$_{8}$ ($\epsilon=$ \scientific{1}{-4})&
    -3655.61091238& 71.43\%\\
    LASSQD$_{9}$ ($\epsilon=$\scientific{1}{-7})&
    -3655.61092764& 73.80\%\\
    LASSQD$_{11}$ ($\epsilon=$\scientific{1}{-8})&
    -3655.61094802& 87.62\%\\
    LASSCI$_{1}$ ($\tau=$\scientific{1}{-1})&
    -3655.53646248& 0.63\%\\
    LASSCI$_{2}$ ($\tau=$\scientific{1}{-2})&
    -3655.57592299& 8.30\%\\
    LASSCI$_{3}$ ($\tau=$\scientific{8}{-3})&
    -3655.59484503& 12.70\%\\
    LASSCI$_{4}$ ($\tau=$ \scientific{7}{-3})&
    -3655.59733211& 20.99\%\\
    LASSCI$_{5}$ ($\tau=$ \scientific{6}{-3})&
    -3655.60045097& 25.54\%\\
    LASSCI$_{6}$ ($\tau=$ \scientific{4}{-3})&
    -3655.60870495& 40.97\%\\
    LASSCI$_{7}$ ($\tau=$ \scientific{3}{-3})&
    -3655.60877031& 41.44\%\\
    LASSCI$_{8}$ ($\tau=$ \scientific{2}{-3})&
    -3655.60988275& 47.62\%\\
    LASSCI$_{9}$ ($\tau=$ \scientific{1}{-3})&
    -3655.61070539& 62.86\%\\
    LASSCI$_{10}$ ($\tau=$ \scientific{1}{-4})&
    -3655.61097614&99.50\%\\

    \hline
    \end{tabular}
    
\end{table}

\nocite{*}

\bibliography{apssamp}
\end{document}